# Spin-polarizing properties of heterostructures with magnetic nano elements*

Armen N. Kocharian[a], Avag S. Sahakyan[b] and Ruzan M. Movsesyan[b]

[a]*California State University, Los Angeles, University Dr., Los Angeles, CA 90032*

[b]*The State Engineering University of Armenia, Teryan St. 105 Yerevan-0009, Armenia*

**Abstract**

The problem of electron resonant and non-resonant scatterings on two magnetized barriers is studied in the one-dimension. The transfer-matrix is built up to *exactly* calculate the coefficient of the electron transmittance through the system of two magnetic barriers with non-collinear magnetizations. The polarization of the transmitted electron wave for resonance and non-resonance transmittances is calculated. The transmittance coefficient and spin polarization can be drastically enhanced and controlled by the angle between the barrier magnetizations

## 1. Introduction

Over the last decade, quantum transport of electrons in semiconductors and heterostructures has been a subject of extensive experimental and theoretical investigations, especially for spin-½ Fermi particle scattering on the strongly magnetized objects (spin-depending scattering potentials). In most cases the spin-dependent potential has a structure which can lead to one dimensional character of electron scattering. The physics of spin-dependent tunneling phenomena has a broad practical applications in spintronics [1] for design of spin manipulated devices such as spin polarized diodes, information processors, quantum computing elements, spin filters, injectors to name just a few. For instance, the tunneling through a non-magnetic semiconductor for filter design was first discussed in Refs. [1,2], where the spin splitting was due to the Rashba interaction [3]. As it is known [4], the most efficient method for observation of electron spin polarization is the resonance tunneling through a system of (multilayers) barriers. In Refs. [5,6] the layer structure is considered to make a *magnetized* quantum well suitable for resonant transmission. This structure provides the two groups of resonance levels for "up-spin" and "down-spin" electrons, respectively, and the transmitted wave becomes strongly polarized due to the carriers resonance transmittance. The spin-dependent tunneling in double magnetic tunnel junctions produced by the impurities in the quantum well can tune the spin polarization and magnetic-resistance [7]. However, as far as the authors are aware, the resonant regime has not been addressed yet neither in non-collinear spintronics or giant magneto-resistance effect. In Ref. [8], a rather simple spin valve transistor-magnetic multilayer is proposed in which the spin-dependent current depends on parallel or antiparallel orientations of magnetization in two magnetic layers.

In the present paper, the electron scattering on a magnetized double potential barrier is considered for non-collinear barrier magnetization vectors with arbitrary angle $\theta$ between magnetic moments in layers. The ferromagnetic element doped into the nonmagnetic semiconductor is commonly used for generation and manipulation of spin-polarized currents. In this spirit the energetic double-barrier profile can correspond to a heterostructure containing two magnetic layers, such as the semiconductor $ZnO$ doped with $Ni$. Its band-gap width, $\Delta = 3,2 - 3,5\,eV$, is large; besides, its Curie temperature and magnetization are large too [9]. The scattering of both spin-polarized and spin-non-polarized electrons is considered and we find that the spin resonant current depends on the angle $\theta$. The spin-polarizing properties of such systems are investigated analytically. In this article the dependences of transmittance coefficient and spin polarization vector of the forward scattered electron wave on the angle $\theta$ are studied in a broad energy range, far from the resonant transmittance and precisely at the resonance. A resonant and non resonant tunneling structure consisting of magnetic tunneling barriers is proposed and the spin-dependent transport properties are analytically calculated. The original multilayer system resembles the behavior of "spin-valve transistor" dependence on angle $\theta$, which prevents current transmission through a system of oppositely polarized magnetic layers

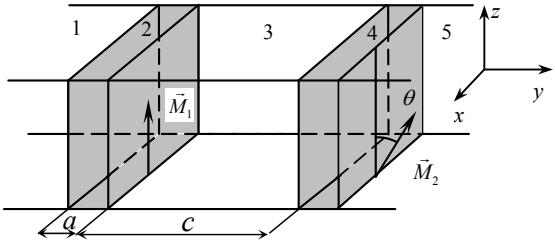

Fig.1. The schematic picture illustrates the five-layer heterostructure (magnetic bilayer with a semiconductor spacer, sandwiched between 1 and 5 metallic layers).. The magnetic moment $\vec{M}_1$ and $\vec{M}_2$ in layers 2 and 4 are localized in xz plane, $M_1$ is parallel to z axis, and the angle between the magnetizations is equal to $\theta$.

An analogous problem was considered in [10], where an electron scattering on a three magnetized δ potential barriers was considered by the Lippman-Shwinger equation. To carry out these calculations we generalized and applied the well-known transfer matrix method developed in Refs. [11-13] for the spin-magnetic interaction.

In general, investigation of electron transfer in magnetic (non-collinear) multilayer systems is a rather complicated problem: the electron scattering on each separate barrier takes place either with a spin flip, or without it, each with a certain probability. Therefore, the number of the intermediate scattering channels increases with the number of barriers. It is apparent that elaboration of analytical methods for calculating the scattering amplitudes of electrons in these systems is very crucial. Taking this into account, one should mention [14], where a Green function of multilayer magnetic systems was built, which makes possible investigation of transport properties of such systems also analytically.

## 2. Theoretical model

Here we propose a simple model consisting of two magnetic (layers) barriers that can be calculated exactly. Let us choose the frame as follows: the $x$ and $z$ axes are in the interface plane and the $y$ axis is perpendicular to the surface (plane) of the barrier shown for a multilayer heterostructure (Fig. 1). We consider the simplest symmetric barriers, and the potential energy of the local electron interaction in the barrier is taken of the following form:

$$H_{int} = \left(V_0 - \gamma\vec{M}\vec{\sigma}\right)\varphi(y), \qquad (1)$$

where $\varphi(y)$ is an even and finite function with carrier $[0,a]$; in the case of rectangular barrier $\varphi(y) = \theta(y)\theta(a-y)$ where $\theta$ is the Heavyside function and $a$ is the width of the barrier. Below, some results are given as examples for the rectangular barrier; nevertheless, the main results (the transmittance coefficients, the polarization vector and polarization degree) are expressed by transmittance and reflectance amplitudes for the single barrier, and one can carry out their qualitative study without specifying the detailed barrier form. Then, we take the energy barrier height equal to $V_0$,

$\vec{\sigma}$ is a vector composed by the Pauli matrices, $\vec{M}$ is the barrier magnetization, $\gamma = g\mu_B/2\mu_0$, $g$ is the g-factor of the electron, $\mu_B$ is the Bohr magneton. In the case of two barriers, one has to add the same function to (1), but transferred by $c$ (the distance between the barriers). The model (1) offers a simple picture for electron transmission and scattering in magnetic semiconductor nanostructures having large forbidden energy gap and magnetic barriers produced by the deposition of ferromagnetic strips.

We take the barrier's vector potential as $\vec{A}(-M_z y, -M_x z, 0)$, $M_z = M\cos\theta$, $M_x = M\sin\theta$, where $\theta$ is the angle between $\vec{M}$ and $z$ axis. It is easy to show that for $\Phi \ll \Phi_0/2\pi$ ($\Phi_0$ is the elementary flux and $\Phi$ is the magnetic flux through the $y$ face of the layer) the orbital part of the Hamiltonian can be neglected. The above inequality can be written as $a_M \gg \sqrt{L_0 L_y}$, where $a_M$ is the magnetic length, $L_0$ is the interface linear size, $L_y$ is its thickness. We can take $L_0 \sim 100 nm$, $L_y \sim 10 nm$ and get $B \ll 6\, Tl$, $(B = \mu_0 M)$, which is obviously satisfied well (analogous estimations one can find in [15]). For the given thicknesses, the commonly used effective mass approximation can be applied [16]. Then, after separation of variables the Schrödinger equation can be reduced to the corresponding one for the envelope wave function:

$$-\frac{\hbar^2}{2}\frac{d}{dy}\left[\frac{1}{m(y)}\frac{d\hat{\psi}}{dy}\right] + H_{int}\hat{\psi} = E\psi, \quad (2)$$

where $E = \varepsilon - \frac{\hbar^2 k_\parallel^2}{2m}$, and $\varepsilon$ is the electron energy, $k_\parallel$ is the electron wave vector component parallel to the interface plane, $m$ is electron effective mass. Below we take $m(y) = m_1$ inside the barrier and equal to $m_2$ outside of it.

### 3. Non-polarized electron wave scattering (a)

We first consider the scattering for the non-polarized electron on the single barrier (1). It must be taken into account that one electron can be always polarized by magnetic field, and, therefore, it is necessary to consider the spin-polarization in the ensemble of electrons. In the absence of electron interaction this implies that the problem of polarization can be solved using the effective single particle approximation. Hence, we take the scattering wave function in the following form:

$$\psi_I = \frac{1}{\sqrt{2}}\left(\hat{I}e^{iky} + \hat{r}e^{-iky}\right), \quad y \leq 0$$
$$\psi_{II} = \frac{1}{\sqrt{2}}\hat{t}e^{iky}, \quad y \geq a, \quad (3)$$

where the column $\hat{I}$ satisfies the condition $\hat{I}^+\hat{I} = 2$. We take it as:

$$\hat{I} = \begin{pmatrix} e^{i\omega} \\ 1 \end{pmatrix}, \quad (3a)$$

where $\omega$ is a random real phase of a uniform distribution. Such choice of $\hat{I}$ is dictated by the condition of non-polarity of the incident electron wave – the electron spin in states (3a) vanishes after averaging with respect to $\omega$, which implies averaging over the electron ensemble. The columns $\hat{t}$ and $\hat{r}$ are the transmittance and reflectance amplitudes, satisfying to the current conservation probability, $\hat{r}^+\hat{r} + \hat{t}^+\hat{t} = 2$. Here the factor $2^{-1/2}$ before $\hat{t}$ and $\hat{r}$ is introduced for convenience. If we take the direction $\vec{M}$ as the axis of quantization, then Schrödinger's equation splits into two independent system of corresponding equations for spinors $\psi_{1,2}$[2]. Solving these equations for the rectangular barrier with suitable boundary conditions [15], we can derive the transmittance and reflectance coefficients for the electrons with opposite spins:

---

[2] *It is easy to show that both the transmittance and reflection coefficients for both single and double identical barriers are independent of the spatial orientation of the magnetization vector. The reason of this is that the incident electron wave is not polarized.*

$$t_\ell = \frac{m_2}{m_1} \cdot \frac{4ikq_\ell e^{-2ika+i\omega}}{\left(q_\ell + i\frac{m_2}{m_1}k\right)^2 e^{-q_\ell a} - \left(q_\ell - i\frac{m_2}{m_1}k\right)^2 e^{q_\ell a}},$$

$$r_\ell = \frac{2\left(q_\ell^2 + \frac{m_2^2}{m_1^2}k^2\right)sh(q_\ell a)}{\left(q_\ell + i\frac{m_2}{m_1}k\right)^2 e^{-q_\ell a} - \left(q_\ell - i\frac{m_2}{m_1}k\right)^2 e^{q_\ell a}}, \quad (4)$$

where

$$k = \frac{1}{\hbar}\sqrt{2m_1 E}, \quad q_\ell = \frac{1}{\hbar}\sqrt{2m_2(V_0 \mp \gamma M - E)}.$$

The effective energy barrier heights for the electrons with opposite spin orientations are $V_0 \pm \gamma M$. Then, it follows from (4) that in the energy region $V_0 < E < V_0 + \gamma M$ and $V_0 - \gamma M < E < V_0$ an over-barrier scattering of electrons takes place, with spins $\vec{s} \uparrow\uparrow \vec{M}$ and $\vec{s} \uparrow\uparrow (-\vec{M})$, respectively.

From (4), after averaging over the ensemble of electrons, transmittance and reflectance coefficients coincide with $|t_\ell|^2$ and $|r_\ell|^2$ respectively. On the other hand, the amplitudes $t_\ell$ and $r_\ell$ correspond to the electrons having spins parallel (anti-parallel) to the barrier magnetizations, since both types of electrons have equal probabilities.

Now we can consider the scattering of non-polarized electrons on embedded two identical magnetic barriers – with equal heights and magnetizations. For this, we need to define the (diagonal) unitary transfer-matrix:

$$\hat{\sigma}\begin{pmatrix}1\\r_1\\1\\r_2\end{pmatrix} = \begin{pmatrix}t_1\\0\\t_2\\0\end{pmatrix}, \quad \hat{\sigma} = \begin{pmatrix}\sigma_1 & 0\\ 0 & \sigma_2\end{pmatrix}, \quad \sigma_\ell = \begin{pmatrix}1/t_\ell^* & -r_\ell^*/t_\ell^*\\ -r_\ell/t_\ell & 1/t_\ell\end{pmatrix}. \quad (5)$$

Then one easily gets the following expression for the transmittance amplitudes,

$$T_{0\ell} = \frac{t_\ell^2 e^{i\omega_\ell}}{1 - r_\ell^2 \exp(2ikc)}, \quad \omega_1 = \omega, \; \omega_2 = 0, \quad (6)$$

and corresponding transmittance coefficient is equal to

$$D = \frac{1}{2}\hat{T}_0^+ \hat{T}_0 = \frac{1}{2}\left(|T_{01}|^2 + |T_{02}|^2\right) = D_\uparrow + D_\downarrow. \quad (7)$$

We define the electron energy spectrum from the following equation:

$$\delta_\ell + kc = \pi(n + 1/2), \quad (8)$$

where $\delta_\ell$ are the forward scattering phases on a single barrier. Then we find that one of the transmittance coefficients equals to 1/2 and the other is exponentially small. In the case of rectangular barriers the phase $\delta_\ell$ in Eq. (8) is defined by the following expression:

$$\delta_\ell = \frac{\pi}{2} - 2ka + \arctan\left\{\frac{2q_\ell k}{\frac{m_2}{m_1}k^2 - \frac{m_1}{m_2}q_\ell^2}cth(q_\ell a)\right\}. \quad (9)$$

### 4. The transfer-matrix

To investigate the electron scattering on the system of barriers with non-collinear magnetizations it is necessary to construct the transfer-matrix for the case when the quantization axis is taken arbitrarily. We redefine transfer-matrix (5) as follows:

$$\tilde{\sigma}\begin{pmatrix}\hat{I}\\ \hat{r}\end{pmatrix} = \begin{pmatrix}\hat{t}\\ 0\end{pmatrix}, \quad \hat{I} = \begin{pmatrix}1\\1\end{pmatrix}, \quad \hat{r} = \begin{pmatrix}r_1\\r_2\end{pmatrix}, \quad \hat{t} = \begin{pmatrix}t_1\\t_2\end{pmatrix}. \quad (10)$$

Then the left and right columns of (5) and (10) are connected through the following matrix:

$$A = \begin{pmatrix}1 & 0 & 0 & 0\\ 0 & 0 & 1 & 0\\ 0 & 1 & 0 & 0\\ 0 & 0 & 0 & 1\end{pmatrix}, \text{ and } A^{-1} = A,$$

and

$$\tilde{\sigma} = A\sigma A, \quad (11)$$

or

$$\tilde{\sigma} = \begin{pmatrix} \alpha^* & -\beta^* \\ -\beta & \alpha \end{pmatrix}, \quad \alpha = \begin{pmatrix} 1/t_1 & 0 \\ 0 & 1/t_2 \end{pmatrix}, \quad \beta = \begin{pmatrix} r_1/t_1 & 0 \\ 0 & r_2/t_2 \end{pmatrix}. \quad (12)$$

If $\vec{M}$ does not coincide with the quantization axis, the Hamiltonian of the spin magnetic interaction, $\gamma \vec{M} \vec{\sigma}$, is non-diagonal and the spinor components in Schrödinger's equation become mixed. We can make the Hamiltonian diagonal by using the unitary transformation $U^+ H U$. We introduce the unitary transformed transmittance $\hat{t}'$ and reflectance $\hat{r}'$ coefficients. It is convenient to transform, $\hat{t}' = U^+ \hat{\tau}$, $\hat{r}' = U^+ \hat{\rho}$, and express $\hat{t}', \hat{r}'$ through $\hat{\tau}$ and $\hat{\rho}$, which are the transmittance and reflectance coefficients of the original problem. Then, using the definition (11), we can obtain

$$\begin{pmatrix} U & 0 \\ 0 & U \end{pmatrix} \tilde{\sigma} \begin{pmatrix} U^+ & 0 \\ 0 & U^+ \end{pmatrix} \begin{pmatrix} \hat{I} \\ \hat{\rho} \end{pmatrix} = \begin{pmatrix} \hat{\tau} \\ 0 \end{pmatrix},$$

and get for the transfer-matrix expression:

$$S = \hat{U}^+ A \sigma A U, \quad (13)$$

where

$$\hat{U} = \begin{pmatrix} U & 0 \\ 0 & U \end{pmatrix}, \quad U = \begin{pmatrix} \cos(\theta/2) & -\sin(\theta/2) \\ \sin(\theta/2) & \cos(\theta/2) \end{pmatrix},$$

and $\theta$ is the angle between the quantization axis and $\vec{M}$.

Using the notations introduced in (12), we can reduce Eq. (13) to

$$S = \begin{pmatrix} U \alpha^* U^+ & -U \beta^* U^+ \\ -U \beta U^+ & U \alpha U^+ \end{pmatrix}. \quad (14)$$

Often, in barrier problems one has to define a transfer-matrix that transforms the barrier's right hand side amplitudes into the left hand side ones. In this case the above-mentioned matrixes are:

$$\tilde{S} = \begin{pmatrix} U \alpha U^+ & U \beta^* U^+ \\ U \beta U^+ & U \alpha^* U^+ \end{pmatrix}. \quad (15)$$

The existence of transfer-matrix and its properties are based on the following fundamental principles: (a) the probability current is conserved; (b) Schrödinger's equation is invariant with respect to time inversion [16]. It is necessary to underline that Eqs. (14) and (15) are derived using these two main principles.

Let us now introduce the following columns:
$$b_I = \begin{pmatrix} I \\ r \end{pmatrix}, \quad b_{II} = \begin{pmatrix} t \\ 0 \end{pmatrix}.$$

Then:
$$S b_I = b_{II}.$$

Schrödinger's equation invariance with respect to time requires complex pairing, magnetization vector inversion and substitution of contra variant spinors for covariant ones, by applying all these simultaneously [16].

We introduce an operator of the magnetization vector inversion: $\hat{K} f(\vec{M}) = f(-\vec{M})$, $\forall f$. Then, $\hat{K} f(-\vec{M}) = f(\vec{M})$, and $\hat{K}^{-1} = \hat{K}$ respectively.

Let us introduce the matrix:
$$\Lambda = \begin{pmatrix} 0 & i\sigma_y \\ i\sigma_y & 0 \end{pmatrix},$$

where $i\sigma_y$ is the metric spinor, transferring from contra variant components to the covariant ones. $[\hat{K}, \hat{\Lambda}] = 0$, because $\hat{K}$ and $\hat{\Lambda}$ work on different variables.

We form a vector-column:
$b_I' = \Lambda K b_I^*$, $b_{II}' = \Lambda K b_{II}^*$. Then $b_{II}' = S(\vec{M}) b_I'$, or $\Lambda K b_{II}^* = S(\vec{M}) \Lambda K b_I^*$. From the other hand, $b_{II}^* = S^*(\vec{M}) b_I^*$, consequently:

$$\Lambda^{-1} S(\vec{M}) \Lambda = S^*(\vec{M}) \quad (16)$$

From the condition of equal probabilities for currents on the left and right of the barrier, we get:

$$J = S^+(\vec{M}) \hat{e} S(\vec{M}), \quad \hat{e} = \begin{pmatrix} e & 0 \\ 0 & -e \end{pmatrix}, \quad (17)$$

where $e$ is the unit $2 \times 2$ matrix.

Conditions (16) and (17) allow reduce the transfer-matrix to the following functional form,

$$S(\vec{M}) = \begin{pmatrix} u_{11}(\vec{M}) & u_{12}(\vec{M}) & u_{13}(\vec{M}) & u_{14}(\vec{M}) \\ -u_{12}^*(-\vec{M}) & u_{11}^*(-\vec{M}) & u_{14}(\vec{M}) & u_{13}(\vec{M}) \\ u_{13}^*(\vec{M}) & u_{14}^*(\vec{M}) & u_{11}^*(\vec{M}) & u_{12}^*(\vec{M}) \\ u_{14}^*(\vec{M}) & u_{13}^*(-\vec{M}) & -u_{12}^*(-\vec{M}) & u_{11}^*(-\vec{M}) \end{pmatrix}, \quad (18)$$

which has the following equivalent structure

$$\begin{pmatrix} A^* & B^* \\ B & A \end{pmatrix},$$

where $A, B$ are $2 \times 2$ quadratic matrixes. Taking into account that the inversion, $\vec{M} \to -\vec{M}$, is equivalent to the substitution $\theta \to \theta + \pi$, it is easy to show that (14) is in agreement with (18).

**5 Non-polarized electron wave scattering (b)**

Here we calculate the transmittance coefficient for the system of two barriers with non-collinear magnetizations. As in section **3**, here we also assume that the wave function of the incident electron is given by (3) and (3a).

One can connect the left hand side (on the barrier) to the right hand side amplitudes using the transfer-matrix (15) relation

$$\begin{pmatrix} \hat{I} \\ \hat{R} \end{pmatrix} = \tilde{S}(0) V(c) \tilde{S}(\theta) V^{-1}(c) \begin{pmatrix} \hat{T} \\ 0 \end{pmatrix}, \quad (19)$$

where $T$ and $R$ are the system's transmittance and reflectance amplitudes, and the quantization axis coincides with the direction of the magnetic field of one of the barriers $(\theta_1 = 0)$, and $V(c)$ is given as follows:

$$V(c) = \begin{pmatrix} e^{-ikc} & 0 \\ 0 & e^{ikc} \end{pmatrix}, \quad e^{-ikc} = \begin{pmatrix} e^{ikc} & 0 \\ 0 & e^{-ikc} \end{pmatrix}. \quad (20)$$

From Eq. (19) we get the matrix equations for amplitudes $T$ and $R$ respectively:

$$T = M^{-1} I = U \left( \alpha U \alpha + \beta^* U \beta e^{2ikc} \right)^{-1} I,$$
$$R = \left( \beta U \alpha + \alpha^* U \beta e^{2ikc} \right) U^+ T. \quad (21)$$

It can be shown that the diagonal elements of $\hat{M}$ are:

$$T_{\uparrow\uparrow} = \frac{1}{\det M} \left( \frac{\cos^2(\theta/2)}{T_{02}} + a \sin^2(\theta/2) \right),$$
$$T_{\downarrow\downarrow} = \frac{1}{\det M} \left( \frac{\cos^2(\theta/2)}{T_{01}} + a \sin^2(\theta/2) \right), \quad (22a)$$

and the anti-diagonal elements are the transmittance amplitudes with turned spin:

$$T_{\uparrow\downarrow} = \frac{1}{2\det M} \sin\theta \left( \frac{1}{T_{02}} - a \right),$$
$$T_{\downarrow\uparrow} = \frac{1}{2\det M} \sin\theta \left( \frac{1}{T_{01}} - a \right), \quad (22b)$$

where

$$\det M = \frac{\cos^2(\theta/2)}{T_{01} T_{02}} + a^2 \sin^2(\theta/2), \quad a = \frac{1}{t_1 t_2} + e^{2ikc} \frac{r_1 r_2^*}{t_1 t_2^*}.$$

To show these, it is sufficient to calculate the matrix elements of $M^{-1}$ between the eigenstates, $\sigma_z$.

It is assumed that the incident electrons on to the left barrier are not spin-polarized and the wave function is given in (3), (3a). Then, after averaging over the ensemble, we have:

$$D = \frac{1}{|\det M|^2} \left\{ (D_\uparrow + D_\downarrow) \cos^2 \frac{\theta}{2} + 2|a|^2 D_\uparrow D_\downarrow \sin^2 \frac{\theta}{2} \right\}. \quad (23)$$

Taking into account that $|T_{01} T_{02} a^2 - 1| \sim o(\gamma M/V_0)$, we can simplify (23) and finally get:

$$D = (D_\uparrow + D_\downarrow) \cos^2 \frac{\theta}{2} + 2\sqrt{D_\uparrow D_\downarrow} \sin^2 \frac{\theta}{2}. \quad (24)$$

The expression Eq. (23), using Eqs. (22a) and (22b), is possible to reduce also to the following form

$$D = \frac{1}{2} \left( |T_{\uparrow\uparrow}|^2 + |T_{\uparrow\downarrow}|^2 + |T_{\downarrow\downarrow}|^2 + |T_{\downarrow\uparrow}|^2 \right). \quad (25)$$

Thus, the transmittance coefficient depends on the angle $\theta$ periodically; its maximum value is reached at $\theta = 0$ $(D_{\max} = D_\uparrow + D_\downarrow)$, and the minimum at $\theta = \pi$ $(D_{\min} = 2\sqrt{D_\uparrow D_\downarrow})$. It is to be noted that $D_{\min}$ is exponentially small even when one of the components is transmitted trough the resonance tunneling. It is seen from (24) that the non-parallel in magnetizations decreases the transmittance coefficient. The dependence on $\theta$ is rather weak in the region far from condition of the resonant transmittance:

$$D \approx (D_\uparrow + D_\downarrow) - 2D_0 \left(\frac{\gamma M}{V_0}\right)^2 \sin^2\frac{\theta}{2}, \quad (26)$$

where $D_0$ is the transmittance coefficient at $M = 0$. At resonance, for instance, when $D_\uparrow = 1/2$ and $D_\downarrow$ is exponentially small, we have:

$$D_{res} = \sqrt{2D_\downarrow} + \frac{1}{2}\cos^2\frac{\theta}{2}. \quad (27)$$

Notice, that the transmittance coefficient dependence on the angle $\theta$ implies similar behavior for the conductance of the system. Note, that as in giant magnetoresistance case the (conductance) transmission is smallest when the magnetic moments in the alternating layers are oppositely aligned and greatest when they are all parallel.

Having the transmittance amplitudes, one can calculate the spin polarization vector of the electrons transmitted through the two barrier system (in $\hbar/2$ units):

$$\vec{P}\overline{\langle \hat{\psi}^+ \hat{\psi} \rangle} = \overline{\langle \hat{\psi}^+ \hat{\vec{\sigma}} \hat{\psi} \rangle}, \quad (28)$$

where the brackets means quantum mechanical averaging and horizontal line implies averaging with respect to the ensemble of the incident electrons, $\psi$ - is the transmittance wave function.

The calculated components of polarization $P_i (i = x, y, z)$, expressed through the transmittance coefficients, $D_{\uparrow,\downarrow}$ using Eq. (24) are the following:

$$P_x = \sin\theta \frac{\left\{(D_\uparrow + D_\downarrow)\cos^2\frac{\theta}{2} - 2\sqrt{D_\uparrow D_\downarrow}\sin^2\frac{\theta}{2} - \sqrt[4]{D_\uparrow D_\downarrow}\left(\sqrt{D_\uparrow}+\sqrt{D_\downarrow}\right)\cos\varphi\cos\theta\right\}}{D}$$

$$P_y = \frac{\sqrt[4]{D_\uparrow D_\downarrow}\left(\sqrt{D_\uparrow}+\sqrt{D_\downarrow}\right)\sin\varphi\sin\theta}{D}, \quad (29)$$

$$P_z = \frac{(D_\uparrow - D_\downarrow)\cos^2\frac{\theta}{2}\cos\theta + \sqrt[4]{D_\uparrow D_\downarrow}\left(\sqrt{D_\uparrow}-\sqrt{D_\downarrow}\right)\cos\varphi\sin^2\theta}{D},$$

where $\varphi = \Delta_1 - \Delta_2$, and here $\Delta_{1,2}$ are the phases of the electrons due to the forward scattering on the two identical magnetic barriers:

$$\Delta_\ell = \arctan\frac{\sin 2\delta_\ell - |r_\ell|^2 \sin 2kc}{\cos 2\delta_\ell + |r_\ell|^2 \cos 2kc} \quad (30)$$

## 6. Scattering of the polarized electron wave

In case, when the incident electron wave is spin-polarized along the magnetization of the first barrier, the spinor $I$ in (21) is to be substituted for one of the eigenstates, $\sigma_z$, for instance, for $\begin{pmatrix}1\\0\end{pmatrix}$. Then, we can get for the transmittance coefficient the following expression

$$D = 2\sqrt{D_\uparrow}\left(\sqrt{D_\uparrow}\cos^2\frac{\theta}{2} + \sqrt{D_\downarrow}\sin^2\frac{\theta}{2}\right) \equiv 2\sqrt{D_\uparrow}d, \quad (31)$$

or

$$D = |T_{\uparrow\uparrow}|^2 + |T_{\uparrow\downarrow}|^2. \quad (32)$$

The calculated transmitted wave polarization vector components are:

$$P_x = \frac{\sin\theta}{d}\left(\sqrt{D_\uparrow}\cos^2\frac{\theta}{2} - \sqrt{D_\downarrow}\sin^2\frac{\theta}{2} - \sqrt[4]{D_\uparrow D_\downarrow}\cos\varphi\cos\theta\right),$$

$$P_y = \frac{\sqrt[4]{D_\uparrow D_\downarrow}\sin\varphi\sin\theta}{d}, \quad (33)$$

$$P_z = \frac{1}{d}\left(\left[\sqrt{D_\uparrow}\cos^2\frac{\theta}{2} - \sqrt{D_\downarrow}\sin^2\frac{\theta}{2}\right]\cos\theta + \sqrt[4]{D_\uparrow D_\downarrow}\cos\varphi\sin^2\theta\right).$$

## 7. Discussion

The polarization degree of the transmitted electron wave is defined as [16]:

$$P(\theta) = \sqrt{P_x^2 + P_y^2 + P_z^2}.$$

For non-polarized incident wave, $P(\theta) \leq 1$, and for fully polarized wave $P(\theta) = 1$. This follows from definition [16], and also from Eqs. (29) – (33). Expressions (29) can be easily analyzed at $\theta = 0$, for the collinear magnetizations. Then the components are

$$P_{0x} = P_{0y} = 0, \quad P_{0z} = \frac{D_\uparrow - D_\downarrow}{D_\uparrow + D_\downarrow}, \quad (34)$$

where $P_{0z}$ is the degree of polarization for the electrons scattered forward.

Far from the resonance, $P_{0z} \approx \gamma M/V_0 \ll 1$, and it is equal to the unit, by neglecting the exponentially small corrections. In the same manner, one can study the situation if the polarizations are collinear and arbitrarily oriented in space. Then we can get for the polarization components the following

$$P_x = P_{oz} \sin\theta, \quad P_y = 0, \quad P_z = P_{oz} \cos\theta. \quad (35)$$

This describes the polarization of the forward scattered wave collinear to the magnetization of the barriers. For the scattering takes place only on one barrier, then

$$P_x = \frac{\gamma M \sin\theta}{V_0}, \quad P_y = 0, \quad P_z = \frac{\gamma M \cos\theta}{V_0}, \quad (36)$$

i.e., this is a case when the polarization can be neglected.

In the case of the non-collinear magnetizations, it follows from the Eqs. (29) that for the non-resonant scattering:

$$P_x \approx (\gamma M/V_0)^2, \quad P_{y,z} \approx \gamma M/V_0, \quad (37)$$

i.e., the polarization is small due to smallness of the parameter $\gamma M/V_0$.

For the resonant scattering, $P(\theta)$ is practically equal to unity, except for the very narrow window near $\theta = \pi$, where there is a narrow dip (Fig. 2). The dependence of $P(\theta)$ on θ follows from Eqs. (29)

$$P(\theta) = \frac{\cos^2(\theta/2)}{2D_{res}}, \quad (38)$$

where $D_{res}$ is defined by formula (27). Qualitatively this result for $P(\theta)$ can be easily understood if we consider a non-polarized unit flux that falls on the left barrier: $I_{0\uparrow} + I_{0\downarrow} = 1$, $I_{0\uparrow} = I_{0\downarrow}$; for the resonant transmission through the first barrier we get, $I_{tr\uparrow}^{(1)} = 1/2$, and $I_{tr\downarrow}^{(1)}$ - is exponentially small, and for the transmission through the second barrier we have

$$I_{tr\uparrow}^{(2)} = \frac{1}{2}\cos^2\frac{\theta}{2}, \quad (39)$$

and again $I_{tr\downarrow}^{(2)}$ is exponentially small. This result (39) is analogous to Malus's law for passing light polarization through crossed polarizers; however, here minimum transmission is obtained when the magnetic moments of the two magnetic films are rotated 180° away from parallel, whereas for the optical case minimum transmission of the two polarizer axes is obtained from a 90° orientation. This is a consequence of the spin's being the source of the magnetization and the $\cos^2(\theta/2)$ dependence, which comes from the spinor transformation when one projects one spin state onto another whose coordinate axis is rotated an angle θ from the first.

For the resonant transmission, $P_{x,z}$ are essentially dependent on $\theta$, but $P_y$ is small, as for the case of the non-resonant transmission: $P_y \sim \gamma M \sin\theta/V_0$.

The space orientation of the polarization of the transmitted wave depends significantly on the angle θ. Note that each point of the curve presented in Fig. 3 corresponds to the end of the polarization vector at each value of $\theta$ for the resonant transmission (the "trajectory" of the polarization vector). Note, that this curve trajectory in space is closed due to the $2\pi$-periodicity of the $P_i$ components.

Another characteristic of the transmitted wave is the orientation of the vector $\vec{P}$ in respect to the z axis: $\Phi(\theta) = \arccos(P_z/P)$ (see Fig. 4). Far from $\theta = \pi$, the dependence of $\Phi(\theta)$ on $\theta$ is close to linear. This dependence was obtained at the resonance of the transmission. In the case of the polarized incident wave, $\Phi(\theta)$ has the meaning of the rotation angle of the polarization vector. Its dependence is also presented on Fig. 4, here the dip vanishes at $\theta = \pi$. Thus the transmission coefficient and the polarization, considered in this article, display a characteristic sharp dip behavior for angles close to anti-parallel orientation of magnetizations in barriers, $\theta = \pi$. We see that the parallel arrangement yield much higher conductance through the barrier than does the antiparallel arrangement. This result is an apparent manifestation of the spin-valve effect [8]. Such strong dependence of magnetoconductance on the non-colinearity angle $\theta$, is also known in non-collinear spintronics for a number of other magnetic multilayer systems. In particular, the spin-transfer has been found also in layered magnetic nanostructures, FNF, which is a contact of normal metallic film of ballistic thickness, sandwiched between two ferromagnetic reservoirs [17].

The two barrier system can strongly polarize the electron wave only at the resonant transmission. Far from the resonance, as well as in the case of one magnetic barrier, the strong polarization effect is practically absent. However, the condition for resonant tunneling through the two barrier system is limited. The effect of resonant transmission can be strongly enhanced in multi barrier system. 1D scattering problem is still applicable for investigations of systems consisting more than two magnetized barriers. To carry out this program for multi barrier system it is necessary to solve the eigenstate and eigenvalue problem for the transfer matrix similar to the one constructed in Sec. 4 for the two magnetic barriers. Thus, we conclude that polarization in the multi-barrier systems can be more effective, since the effect of resonant transmission can strongly enhance spin selectivity of the multilayer.

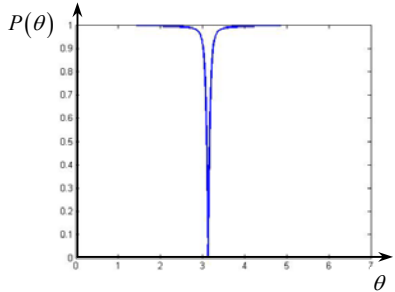

Fig. 2. (Color online) Dependence of the polarization degree of the resonant transmitted wave on the angle $\theta$ for non-polarized incident wave.

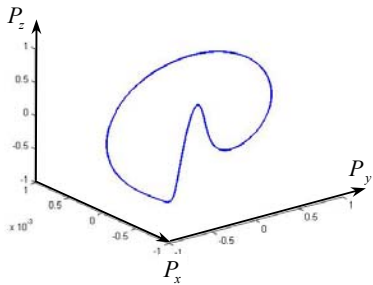

Fig. 3. (Color online) Polarization vector's trajectory for the resonance transmittance for non-polarized incident wave.

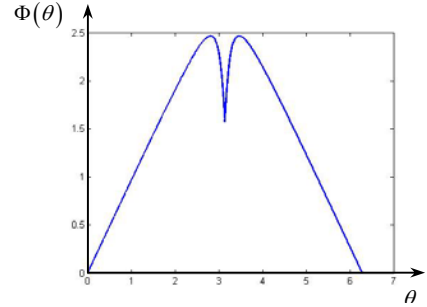

Fig. 4. (Color online) Dependence of $\Phi(\theta)$ on $\theta$ in the case of the non-polarized incident wave, where $\Phi(\theta)$ is the angle between the vector polarization of the resonant transmitted wave and the z axes.

## 8. Conclusion

In conclusion, the transfer-matrix for spin polarization of spin dependent resonant tunnelling in 1D case between two magnetic barriers with different spin orientations is solved exactly. The spin-polarized and spin-non-polarized electron scattering on a magnetized double potential barrier is calculated for various barrier magnetization vectors. Our studies show that transmission of particles through two identical barriers can be found at classically forbidden energy levels formed by these resonant transmissions.

The polarization of an ensemble of incident electrons can be dramatically enhanced due to resonance transmittance, which strongly depends on the relative orientations of magnetic barriers. We find transmittance for non-polarized and partially polarized configurations driven by the relative orientation of magnetization of ferromagnetic barriers. The analytical relations for the angles between polarization and magnetic field are found for non-polarized and fully polarized configurations, which can be changed by an applied magnetic field. The spin-dependent barrier system has also a strong effect on the transmittance, which can be employed for the efficient control of spin polarization via the applied magnetic field. The spin polarized electrons with high efficiency electron transmission probability in magnetic heterostructures can be used in various spin valve devices, magnetic-controlled spin filters, in the field of quantum computing and spin field transistors, [18-20].

Note that the absence of $-e\vec{A}$ term in the Hamiltonian makes the considered problem equivalent to one for the thermal neutron scattering, [21], although the over-barrier transmission energy for thermal neutrons is taking place at the non-resonant region.

### 9. Appendix A

In the non-polarized beam of electrons the spin orientations are random (this corresponds to the case when the half of the spins is oriented along an arbitrary axis and the other half is oriented against the arbitrary chosen axis) and in the incident beam the spin states are a statistical mixture with equal probabilities. For this reason, in general, the spin subsystem cannot be characterized by a wave function, but are described by the density matrix. The density matrix for forward scattering wave is [22]:

$$\rho_f = M^{-1} \rho_{in} (M^{-1})^+, \quad (A1)$$

where $M^{-1}$ is the transmission amplitudes matrix (23), and $\rho_{in} = e/2$ is the density matrix of the non-polarized incident beam. Then, taking into account that:

$$M^{-1} = \begin{pmatrix} T_{\uparrow\uparrow} & T_{\uparrow\downarrow} \\ T_{\downarrow\uparrow} & T_{\downarrow\downarrow} \end{pmatrix}, \quad (A2)$$

where $T_{ij}(i,j=\uparrow,\downarrow)$ are defined by Eqs. (24) and (25). It is easy to obtain the coefficient of transmission through the double barrier using Eq. (A1)

$$D = Sp(\rho_f). \quad (A3)$$

The result (A3) coincides with expression (29), which was obtained through the direct use the wave function (5). In the same way one can obtain that the polarization vector of the forward scattered beam is:

$$\vec{P} = Sp(\vec{\sigma}\rho_f)/Sp(\rho_f), \quad (A4)$$

and it coincides with the expression obtained through Eq. (32). This shows that one can apply wave function (5) of the "non-polarized" electron.

**References**

[1] A. Voskoboinikov, S.S. Lin, C.P.Lee, Phys. Rev. B 58 (1998) 15397.
[2] A. Voskoboinikov, S.S. Lin, C.P. Lee, Phys. Rev. B 59 (1999) 12514.
[3] Y.A. Bychkov, E.J. Rashba, JETP Lett. 39 (1984b) 78.
[4] M.M. Glazov, P.S. Alekseev, M.A. Odnobludov, V.M. Chistiakov, S.A. Tarasenko, J.N. Yascievich, Phys. Rev B. 71 (2005) 155313.
[5] A. Slobodskyy, G. Gould, T. Slobodskyy, Phys. Rev. Lett. 90 (2003) 246601.
[6] X. Jiang, R. Wang, S.van Dijken et al., Phys. Rev. Lett. 90 (2003) 256603.
[7] A. Kalitsov, A. Coho, N. Kioussis, A. Vedyayev, M. Chshiev, A. Granovsky, Phys. Rev. Lett. 93 (2004) 046603.
[8] D.J.Monsma, J.C.Lodder, Th.J.A.Popma, D.Dieny, Phys. Rev. Lett.74 (1995) 5260.
[9] V. Pavle. Radovanovich, Daniel R. Gamelin, Phys. Rev. Lett. 91 (2003) 157202.
[10] Hai-jun hang, Wej-hua Tang, xuean Zhao, Physica B389, (2007) 281.
[11] J. Walker, J. Gathright, Am. Journ. Phys. 62, (1994) 408.
[12] A Zhang, Z. Chao, Q. Shen, X. Dan, Y. Chen. J. Phys. A30 (2000) 5449.
[13] P. Erdos, R. C. Herndon, Adv. Phys. 31 (1982) 65-163.
[14] M.Ye.Zhuravlev, J.D.Burton, A.V.Vedyayev, E.Y. Tsymbal. Journal of Physics A. 38 (2005) 5547.
[15] G. Bastard, Wave Mechanics Applied to Semiconductor Heterostructures, Halsted, New York, 1998.
[16] L.D. Landau, E.M. Lifshitz, in: Quantum Mechanics, Pergamon Press, London, 1977.
[17] A. Brataas, G.E.W. Bauer, P.J.Kelly. Phys. Rep. 427, (2006), 157.
[18] D.P. DiVincenzo, J. Appl. Phys. 85 (1999) 4785.
[19] J. Schliemann, J.C. Egues and D. Loss, Phys. Rev. Lett. 90 (2003) 146801.
[20] T. Matsuyama et al., Phys. Rev. B 65 (2002) 155322.
[21] B.P. Toperverg, Polarized Neutron Scattering, Forschungszentrum Jülich, Series, Mater and Materials, v.12 (2002).
[22] J. R. Taylor, *Scattering theory,* John Willey & Sons Inc., N. Y, 1972.